\documentclass[preprint]{elsarticle}

\usepackage{amsmath}
\usepackage{amssymb}
\usepackage{mathtools}
\usepackage{graphicx}
\usepackage{xcolor}
\usepackage{lineno}
\usepackage[hidelinks]{hyperref}
\usepackage{booktabs}
\usepackage{epstopdf}
\usepackage{float}
\usepackage{xfrac}
\usepackage{natbib}
\usepackage[normalem]{ulem}

\modulolinenumbers[5]

\usepackage[ruled,vlined]{algorithm2e}
\SetKwProg{Timestep}{Time Step}{}{end Time Step}
\SetKwProg{Momentfitting}{Moment Fitting}{}{end Moment Fitting}
\SetKwProg{Boundarymomentfitting}{Boundary Moment Fitting}{}{end Boundary Moment Fitting}

\def\code#1{\mbox{\texttt{#1}}}

\journal{Computer Physics Communications}
\journal{a peer reviewed journal}

\DeclareMathOperator{\sech}{sech}

\newcommand{\muphyii}{\textit{muphyII}}
\newcommand{\muphy}{\textit{muphy}}
\newcommand{\racoon}{\textit{racoon}}

\DeclareMathOperator{\tr}{tr}
\newcommand{\tensor}[1]{\mathrm{#1}}
\renewcommand{\vec}[1]{\mathbf{#1}}
\newcommand{\abs}[1]{|#1|}

\begin{document}

\begin{frontmatter}

\title{The \textit{muphy}II Code: Multiphysics Plasma Simulation on Large HPC Systems}
%\tnotetext[mytitlenote]{Fully documented templates are available in the elsarticle package on \href{http://www.ctan.org/tex-archive/macros/latex/contrib/elsarticle}{CTAN}.}

%% Group authors per affiliation:
%\author{Elsevier\fnref{myfootnote}}
%\address{Radarweg 29, Amsterdam}
%\fntext[myfootnote]{Since 1880.}

%% or include affiliations in footnotes:
\author[rubaddress]{F. Allmann-Rahn}
\author[utexasaddress]{S. Lautenbach}
\author[rubaddress]{M. Deisenhofer}
\author[rubaddress]{R. Grauer}

\address[rubaddress]{Institute for Theoretical Physics I, Ruhr University Bochum, Universitätsstraße 150, 44801 Bochum, Germany}
\address[utexasaddress]{College of Natural Sciences,
The University of Texas at Austin,
120 Inner Campus Dr Stop G2500,
Austin, TX 78712, United States of America}

\begin{abstract}
Collisionless astrophysical and space plasmas cover regions that
typically display a separation of scales that exceeds any code's
capabilities. To help address this problem, the \muphyii{} code
utilizes a hierarchy of models with different inherent scales, unified
in an adaptive framework that allows stand-alone use of models as well
as a model-based dynamic and adaptive domain decomposition. This
requires ensuring excellent conservation properties, careful treatment
of inner-domain model boundaries for interface coupling, and robust
time-stepping algorithms, especially with the use of electron
subcycling. This multi-physics approach is implemented in the
\muphyii{} code, tested on different scenarios of space plasma
magnetic reconnection, and evaluated against space probe data and
higher-fidelity simulation results from literature. Adaptive model
refinement is highlighted in particular, and a hybrid model with
kinetic ions, pressure-tensor fluid electrons, and Maxwell fields is
appraised.
\end{abstract}


\end{frontmatter}

%\linenumbers

\sloppy

\section{Introduction}
\label{sec:introduction}
In collisionless astrophysical and space plasmas, small-scale kinetic effects
can have a significant impact on global dynamics. A widely known example is
magnetic reconnection, where global quantities such as reconnection time,
energy release, and outflow velocities are determined by kinetic effects
associated with a highly non-Gaussian velocity distribution function. On the
other hand, the large-scale structures thus altered will affect the geometry of
the current sheet and, therefore, the reconnection scenario. Numerical modeling of
this coupling of kinetic and global scales is a major challenge in global
simulations of astrophysical and space plasmas. 
Due to the vast separation of scales, very few global kinetic
simulations exist that model the entire plasma region with numerical
schemes based on the Vlasov equation
(e.g.~\cite{peng-markidis-etal:2015}) or a hybrid electron fluid/ion
Vlasov approach
(e.g.~\cite{alfthan-pokhotelov-kempf-etal:2014}). Scales may range, for
example, from the electron inertial length to global scales, such as
the distance of the magnetotail reconnection region to Earth's bow
shock.

Thus, a dynamic multiphysics approach could be the desired strategy. In this
paper, we present such an approach (implemented in the \muphyii{} code) where a hierarchy of
models, ranging from full kinetic Vlasov simulations to five-/ten-moment two-fluid
models, is coupled in a model-adaptive way. This enables us to perform
"intermediate scale" simulations ranging from the electron scales up to several
hundreds of ion inertial lengths. An asymptotic preserving coupling
\cite{degond-deluzet-doyen:2017} to the magnetohydrodynamic equations (MHD) is
also envisioned, which will allow for truly global simulations.

\emph{Multiphysics} techniques and the idea of coupling
different physical models are not new and have been applied in different physical
contexts. Schulze et al. \cite{schulze-smereka-e:2003} couple kinetic
Monte-Carlo and continuum models in the context of epitaxial growth.
Substantial efforts have been undertaken to couple kinetic Boltzmann
descriptions to fluid models (see e.g.~\cite{
tallec-mallinger:1997,
tiwari-klar:1998,
klar-etal:2000,
degond-dimarce-mieussens:2010,
dellacherie:2003,
goudon-jin-etal:2013,
tiwari-etal:2013}.
Kolobov and Arslanbekov \cite{kolobov-arslanbekov:2012} described the
transition from neutral gas models to models of weakly ionized plasmas.

In the context of space plasma physics, Sugiyama and Kusano \cite{sugiyama-kusano:2007} have begun coupling MHD and particle-in-cell (PIC) models and tested this approach on a 1D Alfv\'en wave test problem where the MHD and PIC regions were fixed in advance.
Markidis et al. \cite{markidis-henri-etal:2014} introduce a bulk coupling of a
four-moment two-fluid/Maxwell solver with an embedded PIC code. The PIC code is
used to calculate the closure for the stress and pressure tensors. To achieve
the large time steps compatible with the MHD description, an implicit Maxwell
solver is used.
Daldorff et al. \cite{daldorff-etal:2014} introduced interface coupling to
combine kinetic PIC simulations (using the semi-implicit iPIC3D code) with Hall MHD
fluid codes (using the BATS-R-US framework).
A closely related approach is presented in \cite{makwana-keppens-lapenta:2017}
using MPI-AMRVAC as the MHD solver.

This strategy was applied to global simulations of the magnetotail
\cite{walker-lapenta-etal:2019}, with the crucial modification of only one-way
coupling of the MHD with the PIC solver. The feedback effect of the PIC
simulation on the MHD fluid quantities was neglected.

Two-way and adaptive coupling strategies with applications to Magnetosphere simulations
were presented in \cite{wang-chen-toth:2022a, wang-chen-toth:2022b, shou-tenishev-chen-toth:2021}.
A recent summary of kinetic modeling/MHD modeling of the magnetosphere can be
found in \cite{markidis-olshevskyetal:2021}.
Also worth mentioning is the multi-level, multi-domain method for
Particle-In-Cell simulations \cite{innocenti-lapenta-etal:2013}, which bypasses
the reshaping of weight functions in adaptive PIC simulations.
\textit{Implicit} time integration is important in these treatments, such that
fast waves, e.g.~light, Langmuir, Whistler, and electron drift waves, are
damped. Following this approach, an asymptotic preserving scheme to the quasineutral could
be achieved as described in \cite{degond-deluzet-doyen:2017}.

Concerning the coupling to the large-scale MHD equations, Ho, Datta, and Shumlak
\cite{ho-datta-shumlak:2018} presented a detailed study and numerical tests
on the coupling of the five-moment fluid/Maxwell and the MHD model.
They discuss the issue of interface fluxes in the transition from the MHD to
the two-fluid model. The difficulty lies in the lack of information on electron
quantities. Two approaches have been presented to obtain this missing
information. The first leads to conservative fluxes, while the second is
physically consistent with the MHD assumptions. The interesting observation was
that conservative coupling leads to oscillations, while physically consistent
but not conservative coupling leads to a stable scheme.

Another strategy aimed at global simulation is to combine PIC or hybrid
simulations with adaptive mesh refinement (see e.g.~\cite{fujimoto:2018},
\cite{papadakis-pfau-kempf-etal:2022}).

The development of the {\em mu}lti-{\em phy}sics plasma code \muphy{} started with the implementation of a semi-Lagrangian Vlasov code
\cite{schmitz-grauer:2006-1}.
For the Lagrangian treatment, we developed the backsubstitution method
\cite{schmitz-grauer:2006-2}
in accordance with Darwin's approach to Maxwell's equation to avoid
the propagation of light waves
\cite{schmitz-grauer:2006-3}.

The first version of the \muphy{} framework was presented in
\cite{rieke-trost-grauer:2015}. It allowed a static coupling of a kinetic
Vlasov code with a two fluid/five-moment model. It utilized distributed MPI/CUDA
programming to run on our local cluster of 64 Tesla S1070 GPUs.

In the next development step \cite{lautenbach-grauer:2018}, we aimed at
dynamic adaptive coupling of different models (e.g.~five-moment, ten-moment,
kinetic Vlasov description) where the choice of the model depends on physically
motivated model-refinement criteria. This implementation served as a prototype
implementation that was not optimized for production runs.

The current version of \muphy{} presented in this paper, \muphyii{}, is a complete rewrite of
the prototype implementation aimed at high performance on the JUWELS Booster
GPU cluster at the Forschungszentrum Jülich \cite{juwels}. In addition, it contains a new
strategy to conserve energy in the Vlasov simulations as well as subcycling
methods for the fast electron dynamics and Maxwell's equations.

\muphyii{} has already been used extensively in studies of two-fluid simulations
of three-dimensional reconnection to understand the role of the lower-hybrid
drift instability \cite{allmann-rahn-lautenbach-grauer-etal:2021}. In
\cite{allmann-rahn-lautenbach-grauer:2022} we introduced an energy-conserving
Vlasov solver. The bulk coupling to the ten-moment/two-fluid equations allows
the choice of coarse velocity resolutions. The resulting scheme was used to
study MMS reconnection events. In \cite{allmann-rahn-grauer-kormann:2021}, we
implemented a Lagrangian Vlasov solver where the velocity space was
approximated by a low-rank hierarchical Tucker decomposition. The low-rank
scheme was able to reduce the number of degrees of freedom by almost two orders
of magnitude.

\section{Physical Models}
\label{sec:models}

In \muphyii{}, collisionless models with different levels of physical accuracy and computational cost
are implemented. The exact kinetic model for collisionless plasmas, which takes the complete
velocity space information into account, describes the plasma using
a distribution function $f_s(\mathbf x, \mathbf v, t)$ for each particle species $s$. The evolution
of the distribution function in position space $\mathbf x$, velocity space $\mathbf v$ and time $t$ is given by the Vlasov equation
\begin{equation}
\frac{\partial f_{s}}{\partial t} +\mathbf{v} \cdot \nabla_x f_{s} + \frac{q_{s}}{m_{s}} (\mathbf{E + v \times B}) \cdot \nabla_{v} f_{s} = 0,
\label{eq:vlasov} \end{equation}
which depends on the electric field $\mathbf E$, magnetic field $\mathbf B$, and particle species charge and mass, $q_s$ and $m_s$.
We will refer to \eqref{eq:vlasov} as \emph{Vlasov model}.

Integrating over velocity space and truncating the hierarchy after the pressure equation
yields a reduced set of equations for
particle density $n_s = \int f_S(\mathbf x, \mathbf v, t) \textup d\mathbf v$,
mean velocity $\mathbf u_s= \sfrac{1}{n_s} \int \mathbf v f_s(\mathbf x, \mathbf v, t) \textup d\mathbf v$,
and momentum flux density $\mathcal P_s = m_s \int \mathbf v \mathbf v f_s(\mathbf x, \mathbf v, t) \textup d\mathbf v$. They constitute
the ten-moment multi-fluid equations,
\begin{equation}
\frac{\partial n_{s}}{\partial t} + \nabla \cdot (n_{s} \mathbf{u}_{s}) = 0,
\label{eq:tenmoment_continuity} \end{equation}
\begin{equation}
m_{s} \frac{\partial (n_{s} \mathbf{u}_{s}) }{\partial t} 
- n_{s} q_{s} (\mathbf{E} + \mathbf{u}_{s} \times \mathbf{B}) + \nabla \cdot \mathcal{P}_{s} = 0,
\label{eq:tenmoment_movement} \end{equation}
\begin{equation}
\frac{\partial \mathcal{P}_{s}}{\partial t}
- q_{s} (n_{s} \text{sym}(\mathbf{u}_s \mathbf{E})
+ \frac{1}{m_{s}} \text{sym}(\mathcal{P}_{s} \times \mathbf{B}))
+ \nabla \cdot \mathcal{Q}_{s} = 0.
\label{eq:tenmoment_energy} \end{equation}
Here, the standard product is the outer (tensor) product, $(\mathbf a \mathbf b)_{ij} = a_ib_j$, the vector product has been generalized to tensors, $(\mathcal A\times\mathbf b)^{ijk} = \epsilon_{klm}A^{ijl}b^m$, and the symmetric operator sums over all cyclic permutations of uncontracted indices, $\mathrm{sym}(\mathcal A)_{ijk} = A_{ijk}+A_{jki}+A_{kij}$.

We close this set of equations with a Landau-based model for the heat flux
$\mathrm Q_s = \int \mathbf v \mathbf v \mathbf v f_s(\mathbf x, \mathbf v, t) \textup d\mathbf v = \mathcal Q_s - \mathrm{sym}(\mathbf u_s \mathcal P_s) + 2 m_s n_s \mathbf u_s \mathbf u_s \mathbf u_s$
that receives information from the gradient of the temperature $\mathrm T_s = (\mathcal P_s - mn \mathbf u_s \mathbf u_s)/(n_s k_{\textup B})$,
\cite{allmann-rahn-lautenbach-grauer:2022,allmann-rahn-trost-grauer:2018,allmann-rahn-lautenbach-grauer-etal:2021}
\begin{equation}
\nabla \cdot \mathrm{Q}_{s} = -\frac{\chi}{k_{s,0}}\,n_s\,v_{\textup{th},s}\,\nabla^{2}\,\mathrm{T}_{s}\,,
\label{eq:tenmoment_closure}\end{equation}
with thermal velocity $v_{\textup{th},s} = \sqrt{2T_s/m_s}$ and a free parameter $\chi/k_{s,0}$.
We will refer to the set of equations (\ref{eq:tenmoment_continuity},\ref{eq:tenmoment_movement},\ref{eq:tenmoment_energy},\ref{eq:tenmoment_closure}) as \emph{ten}-moment fluid model.

By assuming isotropic pressure and zero heat flux,
tensorial pressure reduces to the energy density scalar $\mathcal E_s=\sfrac{m_s}{2} \int v^2 f_s(\mathbf x, \mathbf v, t) \textup d\mathbf v$, resulting in a total of
five fluid equations,
\begin{equation}
\frac{\partial n_{s}}{\partial t} + \nabla \cdot (n_{s} \mathbf{u}_{s}) = 0,
\label{eq:fivemoment_continuity} \end{equation}
\begin{equation}
m_{s} \frac{\partial (n_{s} \mathbf{u}_{s}) }{\partial t}
- n_{s} q_{s} (\mathbf{E} + \mathbf{u}_{s} \times \mathbf{B})
+ \frac{1}{N} \nabla (2\mathcal{E}_{s} - m_s n_s \mathbf{u}_s^2)
+ \nabla \cdot (m_s n_s \mathbf{u}_s  \mathbf{u}_s) = 0,
\label{eq:fivemoment_movement} \end{equation}
\begin{equation}
\frac{\partial \mathcal{E}_{s}}{\partial t}
- q_s n_s \mathbf{u}_s \cdot \mathbf{E}
+ \frac{1}{N} \nabla \cdot \Big(\mathbf{u}_s ((N+2) \mathcal{E}_s - m_s n_s u_s^2)\Big) = 0.
\label{eq:fivemoment_energy} \end{equation}
where $N$ is the dimensionality of velocity space.
We will refer to the set of equations (\ref{eq:fivemoment_continuity},\ref{eq:fivemoment_movement},\ref{eq:fivemoment_energy}) as \emph{five}-moment fluid model.

The evolution of electric and magnetic fields $\mathbf E$ and $\mathbf B$ is determined by Maxwell's equations
\begin{equation}
\nabla \cdot \mathbf{E} = \frac{\rho}{\epsilon_{0}}\,,\quad
\nabla \cdot \mathbf{B} = 0\,,\quad
\nabla \times \mathbf{E} = - \frac{\partial \mathbf{B}}{\partial t},\quad
\nabla \times \mathbf{B} = \mu_{0} \mathbf{j} + \mu_{0} \varepsilon_{0} \frac{\partial \mathbf{E}}{\partial t}\,
\label{eq:maxwell} \end{equation}
which use charge density $\rho = \sum_s q_s n_s$, current density $\mathbf j = \sum_s q_s n_s \mathbf u_s$, vacuum permittivity $\varepsilon_0$, and vacuum permeability $\mu_0$.
In the electrostatic case, the electric field $\mathbf E$ is given by Poisson's equation
\begin{equation}
\nabla^2 \Phi = -\frac{\rho}{\epsilon_0}
\end{equation}
where the electric field is related to the electric potential $\Phi$ through $\mathbf{E} = -\nabla \Phi$.

Currently, \muphyii\, uses the same mesh spacing for all physical models. The coupling of the five-moment model with our adaptive mesh code \racoon\, \cite{dreher-grauer:2005} is still in progress.

\section{Code Design and Performance}
\label{sec:code_design}

The {\em 2nd}-generation {\em mu}lti-{\em phy}sics plasma framework, \muphyii{}, is written in C++ and Fortran, and utilizes the MPI standard for distributed memory parallelism.
All solvers are fully three-dimensional and GPU-accelerated using the OpenACC API.
The code was designed to prioritize performance, maintainability, and flexibility,
and we want to elaborate in this section on the design choices.

C++ was chosen for the framework and logic parts of the code because of its
flexibility. For the pure numeric calculations, Fortran was chosen because of
the simplicity of its built-in multi-dimensional array operations. This allows
for excellent compiler optimization and performance out of the box and makes
the numeric implementation easy to write and read at the same time. Good
readability and maintainability are very important for a scientific code because
(i) it must be easy to check whether a numerical method is implemented
correctly, and (ii) the personnel working with the code fluctuates often,
especially in university research. In particular, for students who are new to code
development and do not only apply the code for physical research but also do
research on new numerical methods, a simple code design is crucial.
Conveniently, this often also leads to good performance. However, new physical
models (like relativistic fluid or kinetic descriptions) can be added in any
language like C++ or Julia. They only have to comply with the C++ interface
functions.

% parallelization
The distributed memory parallelization utilizes MPI and a classical domain
decomposition approach. The physical domain is divided into multiple
subdomains (we call them ``blocks''), which are each handled by separate
computing processes and can handle a different type of the underlying physics
(e.g. Vlasov model for ions and ten-moment fluid model for electrons coupled
through Maxwell's equations). The exchange of information between neighboring
subdomains is done via MPI and makes use of boundary cells. Inside each subdomain, 
shared memory parallelization is available and implemented with OpenACC so that 
GPUs can be utilized. CUDA-aware MPI is supported for fast inter-node communication.

% OpenACC
The choice of favoring OpenACC over low-level implementations like CUDA has been made for multiple reasons. The use of OpenACC significantly
reduces development time and improves code readability and maintainability. The
optimized OpenACC code performs well without rigorous adaption to certain device architectures.
Furthermore, it can be run on accelerators from different manufacturers as well as multi-core CPU systems\,
which makes the \muphyii{} framework very flexible.
Finally, the same code base can be used for GPU- and CPU-based supercomputers, again increasing
flexibility and maintainability. For good performance on GPUs, it is necessary to manage
the GPU data transfers explicitly. Concerning the compute kernels, i.e.\ operations that are
carried out on the GPU, there are two options available in OpenACC:
The \code{kernel} construct leaves the compiler complete freedom for optimization, whereas the
\code{parallel} construct gives the developer some possibilities for manual optimization.
A very effective optimization strategy that is often used in \muphyii{} is to collapse nested loops over
different array dimensions into two loops: One loop that utilizes \code{gang} parallelization
and one that utilizes \code{vector} parallelization, where OpenACC gangs can be compared to
CUDA thread blocks and vectors to CUDA threads within the thread block.

\begin{figure}[h!]
    \centering \includegraphics[width=0.85\textwidth]{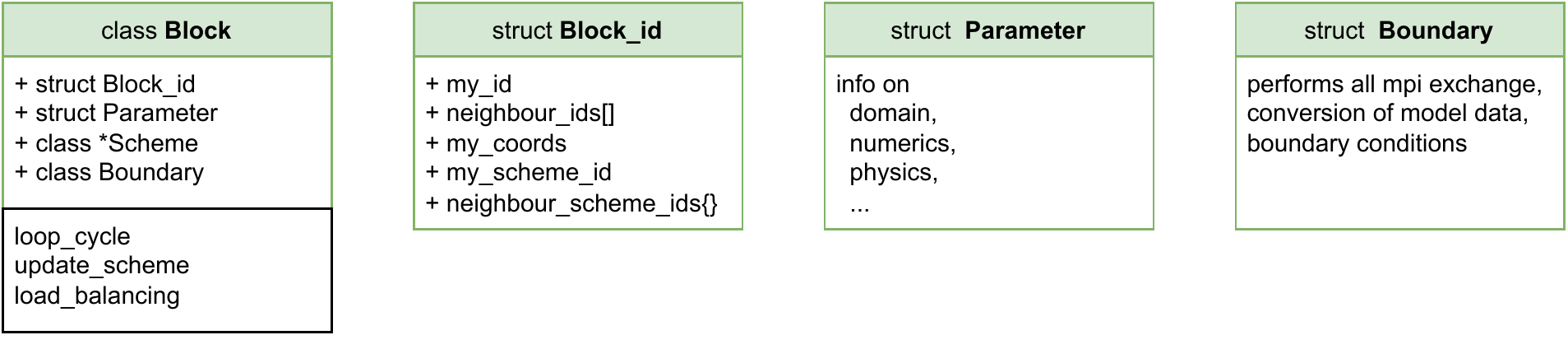}
    \caption{The framework of \muphyii{}, which is responsible for the parallelization, 
             boundary exchanges and conversions, data output and scheme/model adaption is mostly hidden from the user.}
    \label{fig:framework}
\end{figure}

\begin{figure}[h!]
    \centering \includegraphics[width=0.85\textwidth]{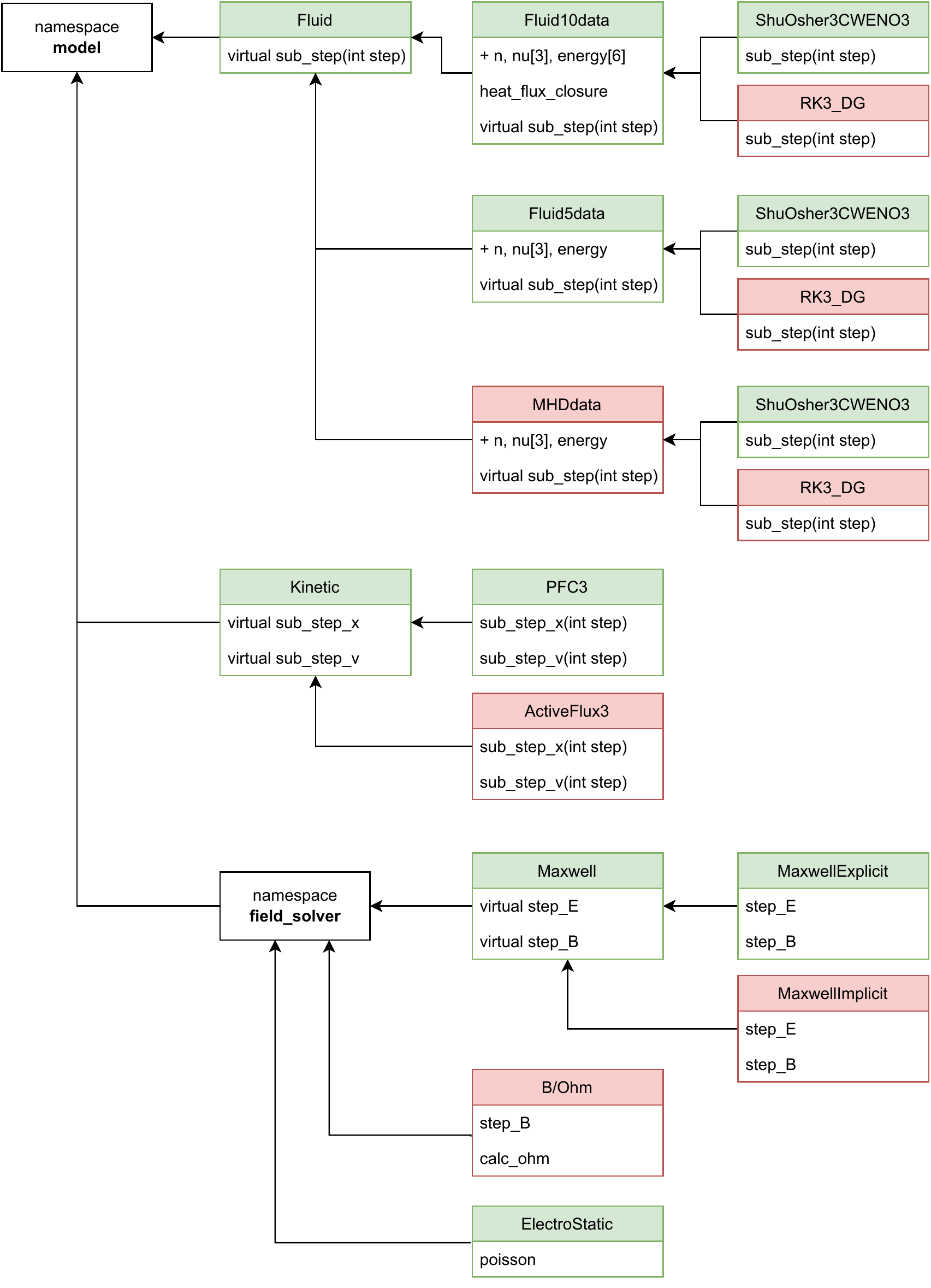}
    \caption{The class Scheme contains all possible combinations of the Model classes. 
             Schemes and models should be easily adjustable and added by the user.
             Models colored green are stable, whereas the models colored red are work in progress.}
    \label{fig:models}
\end{figure}

% layout
The central object in the program layout of \muphyii{} is a block that is associated
with an MPI process. Each block holds a ``scheme'' object that implements the order
of numerical solver function calls and boundary exchange calls. The scheme object
holds multiple ``model'' objects which contain the model data (for example the distribution
function in a Vlasov equation model or the electromagnetic fields in a Maxwell's equations model)
as well as the implementation of the numerical solvers. An example of a scheme would be a \{Vlasov electrons, Vlasov ions,
Maxwell\} scheme, which has a Vlasov model object for each of the species and a Maxwell model
object. The design philosophy where each block/subdomain in the domain decomposition can hold
a different physical scheme meshes well with the multiphysics approach of \muphyii{}.
Depending on the physical necessity, the subdomains can be treated with different physical
models. Simplified UML diagrams summarizing these features are shown in figures \ref{fig:framework} and \ref{fig:models}.
%
% load balancing
In such spatially coupled simulations, the blocks with kinetic models take
significantly more computational time than those with fluid models. The difference
in computational cost is addressed using the CUDA multi-process service (MPS). There,
multiple MPI processes can share a single GPU. The idea is to have one or few
kinetic blocks on each GPU, together with a large number of fluid blocks. Then, ideally, the
few kinetic blocks use the GPU to full capacity in the first part of the numerical scheme,
and the many fluid blocks together also use the GPU to full capacity in the second
part of the numerical scheme. Load balancing is implemented such that equally many
expensive and less expensive models are computed by each GPU.

% scaling
\begin{figure}
\centering \includegraphics[width=0.85\textwidth]{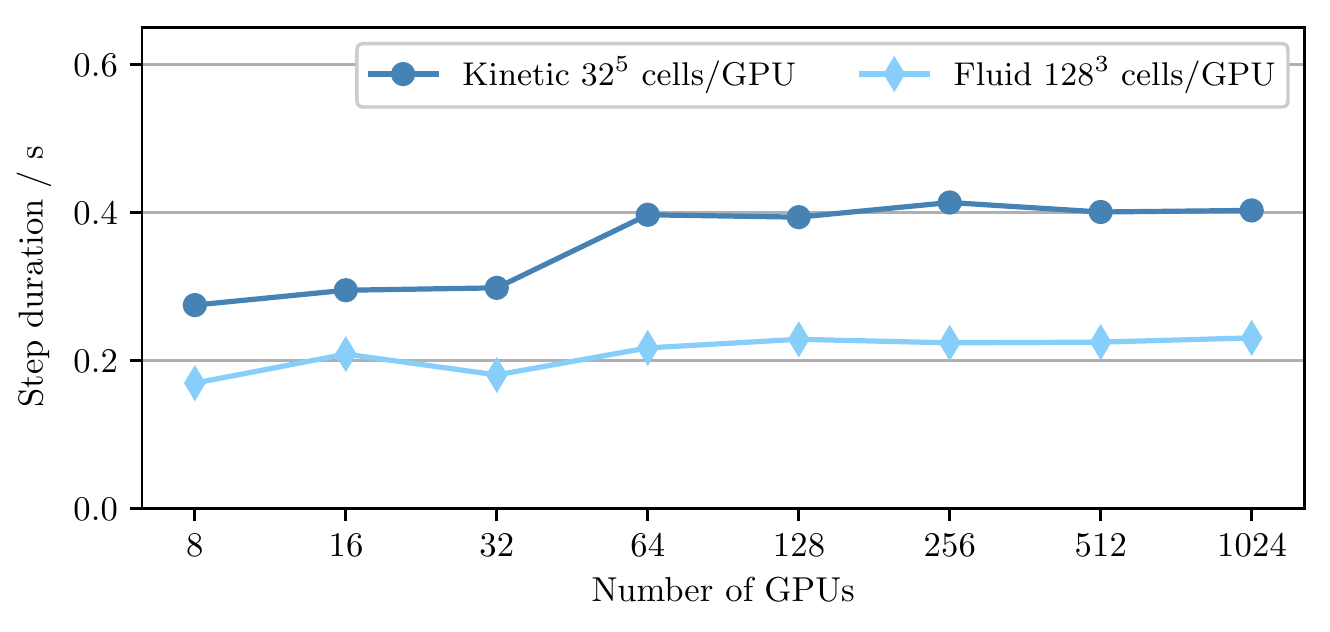}
\caption{Computing time scaling of \muphyii{} on the JUWELS-booster supercomputer.
Shown is the weak scaling where the number of cells and the number of GPUs is increased at
the same time.}
\label{fig:muphy2_scaling}
\end{figure}

The weak scaling of \muphyii{} on the JUWELS-booster supercomputer \cite{jsc-juwels:2021} is shown in
fig.~\ref{fig:muphy2_scaling}. It scales excellently up to 1024 A100 GPUs (more
than 7 million GPU cores), and the scaling is expected to continue in the same
manner towards higher numbers. The increase in step duration
at 64 GPUs is due to the network layout of JUWELS-booster. In JUWELS-booster,
network cells consisting of 48 nodes are connected with very high bandwidth,
and multiple network cells are connected with slightly lower bandwidth.
The increase in step duration is caused by communication over multiple
network cells and is unrelated to the implementation. The scaling study was
performed in double precision, making use of the matched moment Vlasov method 
(see sec.~\ref{sec:solvers}) in the kinetic case and the ten-moment multi-fluid 
model (see sec.~\ref{sec:solvers}) with temperature gradient heat flux closure 
\eqref{eq:tenmoment_closure} in the fluid case, and using 15 subcycles 
(see sec.~\ref{sec:electron_subcycling}) of the FDTD Maxwell solver 
(see sec.~\ref{sec:solvers}).
As a physical setup, we chose an Orszag-Tang vortex for the kinetic solver
in 2D3V \cite{groselj-cerri-navarro:2017} and for the fluid solver in 3D \cite{groselj-mallet-loureiro:2018}.
The time step duration was averaged over 100 steps after a warm-up phase of 50 steps.

\section{Numerical Methods}
\label{sec:numerical_methods}

  \subsection{Solvers}
  \label{sec:solvers}
  
For the solution of the physical equations given in Sec.~\ref{sec:models}, different
numerical solvers are implemented in \muphyii{}. The Vlasov equation \eqref{eq:vlasov} is
solved on a phase space grid with a semi-Lagrangian scheme \cite{cheng-knorr:1976} using
the positive and flux-conservative (PFC) method \cite{filbet-sonnendruecker-bertrand:2001}.
We implement the spatially third-order version and use the limiter only for small values of the phase space distribution function to guarantee positivity.
The position space and velocity space updates are performed with leapfrog-like Strang
splitting.
Formally, the Strang method is only a second-order method with respect to time. 
However, the special nature of the advection terms in the Vlasov equation leads empirically
to a method of higher than second order. As a result, a time order of $2.4$ is usually obtained.
We perform further splitting of the advection operator in the three spatial directions
and the three velocity space directions, respectively. The splitting in the three spatial directions introduces no additional errors since the advection operators commute.
The velocity space splitting
is realized via backsubstitution \cite{schmitz-grauer:2006-2}. To conserve
total energy and to relax the numerical requirements concerning velocity space resolution,
the matched-moments Vlasov solver \cite{allmann-rahn-lautenbach-grauer:2022}
is utilized (see Algorithm \ref{alg:timestep} for details).

The ten- and five-moment equations are solved by means of the CWENO finite volume
method \cite{kurganov-levy:2000}. Time integration is done with a three-step
Runge-Kutta method with good stability properties \cite{shu-osher:1988}.
Several heat flux closures to the ten-moment
equations are implemented: The pressure gradient closure \cite{allmann-rahn-trost-grauer:2018,allmann-rahn-lautenbach-grauer-etal:2021},
the temperature gradient closure \cite{allmann-rahn-lautenbach-grauer:2022},
the temperature gradient closure by Ng et al.\ (2020) \cite{ng-hakim-wang-etal:2020}
and the isotropization closure by Wang et al.\ (2015) \cite{wang-hakim-bhattacharjee-etal:2015}.
The gradient closures are subcycled, as they often need a lower time step for stability
than the fluid equations themselves.
For numerical stability, an explicit floor is set for the pressure.

Maxwell's equations are solved via the finite difference time domain (FDTD) method.
The two Maxwell's equations that include time derivatives are evolved, and it is, by
construction of the method, ensured that $\nabla \cdot \mathbf{B} = 0$. Furthermore,
Gauss's law $\nabla \cdot \mathbf{E} = \rho / \epsilon_0$ is fulfilled in a vacuum or
when the numerical method that provides the current density meets some requirements.
For good preservation of Gauss's law, the current density in the matched-moments Vlasov and fluid
schemes is calculated from the CWENO cell fluxes.
For electrostatic simulations, a simple iterative Poisson solver is implemented
that is parallelized via additive Schwarz iterations.

  \subsection{Time Stepping Scheme}
  \label{sec:time_stepping}

\begin{figure}
\centering \includegraphics[width=0.65\textwidth]{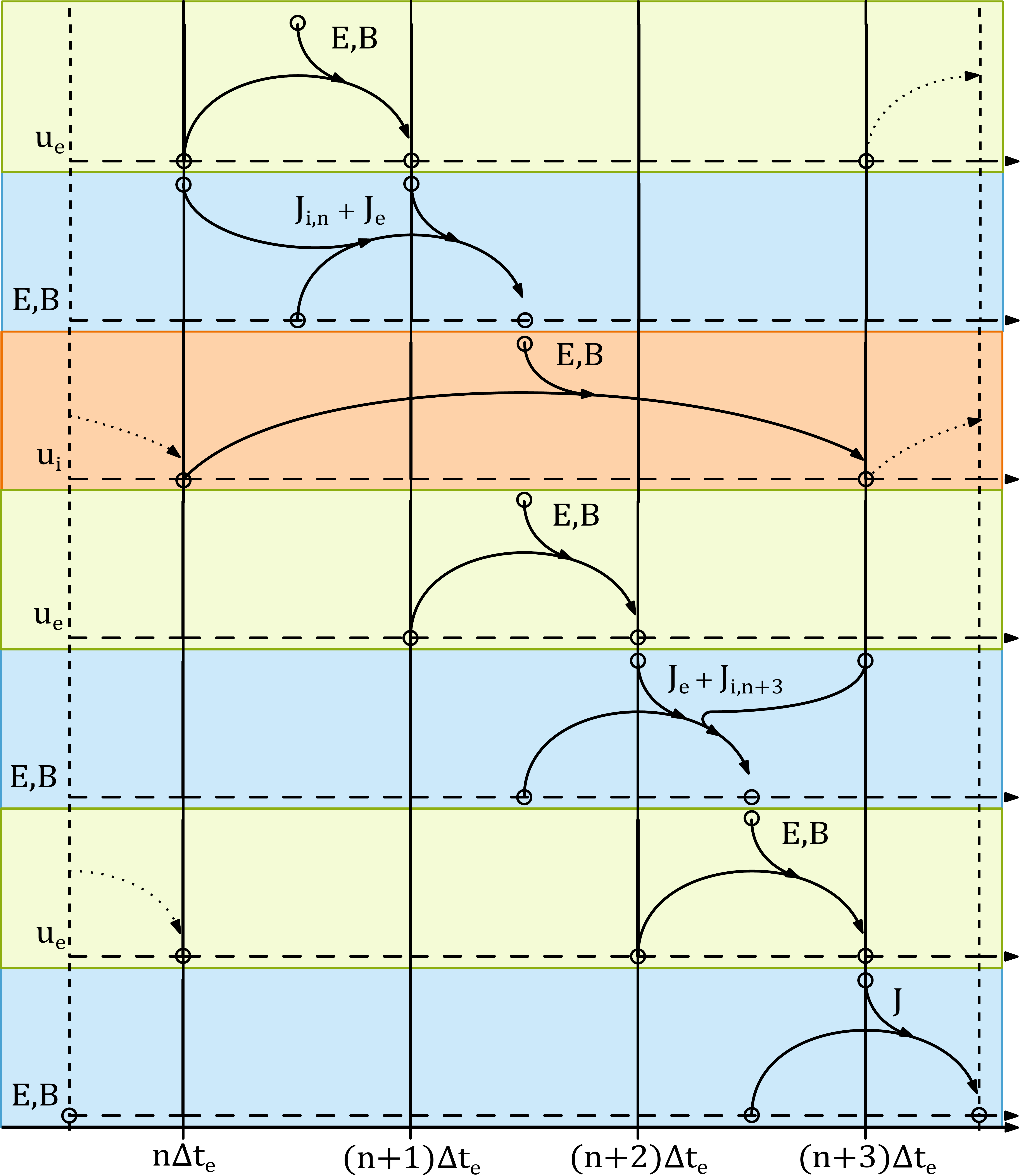}
\caption{Schematic diagram of the time stepping with electron subcycling for $N_e = 3$. Operations are executed from top to bottom, and numeric time goes left to right. Advance operations are highlighted with colors based on their context for visual aid (blue: fields solver, green: electron plasma solver, orange: ion plasm}a solver).
\label{fig:step_subcycling}\end{figure}

\begin{figure}
\centering \includegraphics[width=0.65\textwidth]{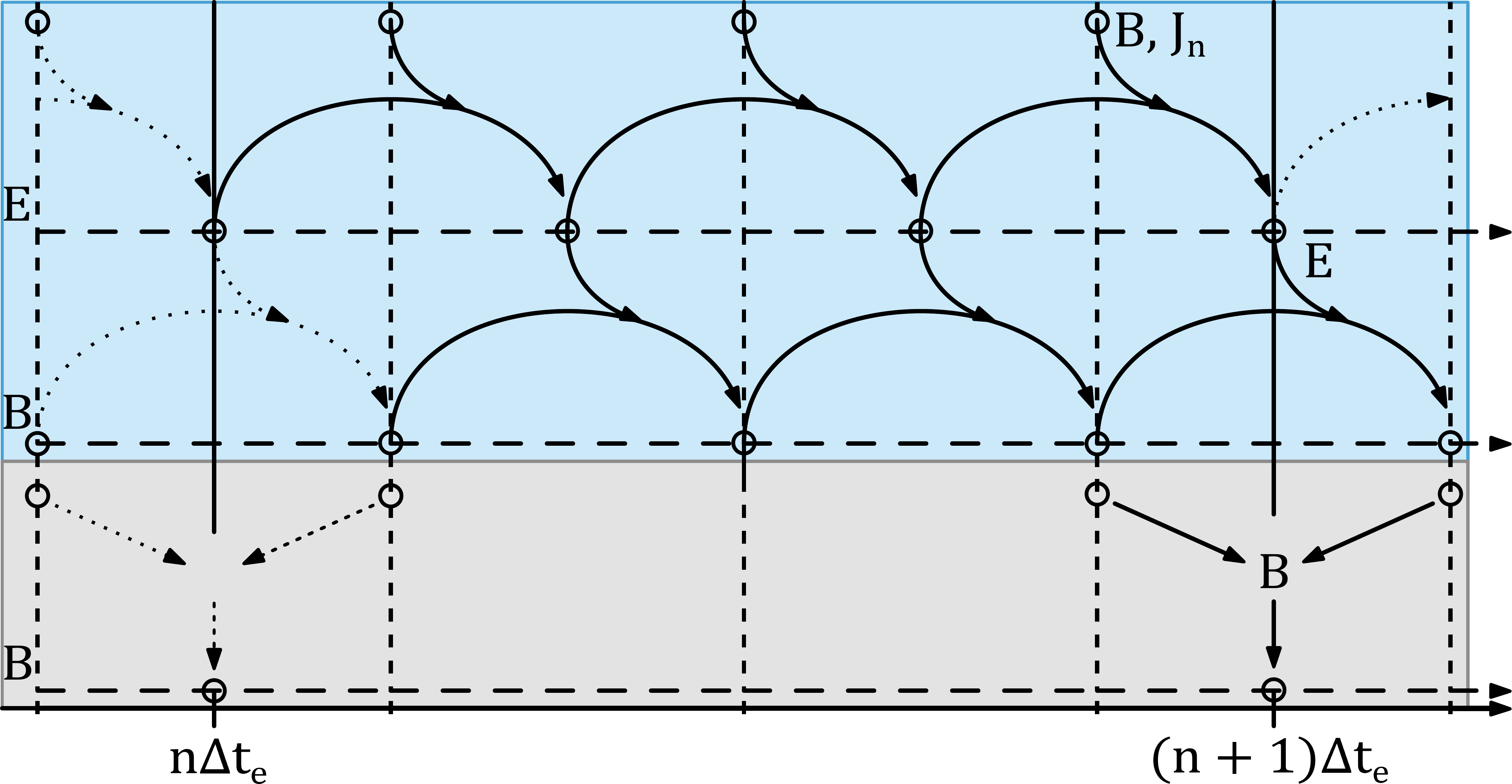}
\caption{Schematic diagram of the leapfrog-like Maxwell subcycling for $N_m = 3$. Operations are executed from top to bottom; numeric time goes left to right. Operations are highlighted with colors based on their algorithmic context for visual aid (blue: advance, grey: interpolation).}
\label{fig:maxwell_subcycling}\end{figure}

To take advantage of the different time resolution requirements of Maxwell's equations and the two particle species' equations, the models have the option to employ subcycling to use different step sizes for each solver.
The much smaller mass of electrons compared to ions typically leads to much faster motion and requires smaller steps, both for fidelity and stability.
For non-relativistic particles, the speed of light is usually also much higher, even when using reduced values.
To reflect this, the subcycling is arranged into two parts.
For every electron step, the Maxwell update is split into $N_m$ Maxwell substeps, and every ion step corresponds to $N_e$ electron substeps.
The electron and Maxwell states are integrated using Strang splitting, which offsets the latter by half the electron time step.

The Runge-Kutta integrator used in the ten and five-moment solvers, and consequently also the ten-moment solver in the moment-matching algorithm, works with Maxwell field data provided at $t + \Delta t_s/2$.
For the case of no electron subcycling, the offset due to Strang splitting ensures that this data is calculated for the right times.
With more than one electron cycle, this is no longer necessarily true for the ion step.
Choosing an uneven amount of electron substeps allows computing the ion step without any additional interpolation by rearranging it after the $\lfloor N_e / 2 \rfloor$-th Maxwell step, as seen in Fig. \ref{fig:step_subcycling}.
The Maxwell solver itself uses Strang splitting between $\mathbf{E}$ and $\mathbf{B}$ and the linear interpolation used to obtain $\mathbf{B}$ at the full Electron step times only needs to be calculated at the last substep (see Fig. \ref{fig:maxwell_subcycling}).

While $N_m$ is manually set as a parameter of the simulation, $N_e$ and the step sizes for each solver are chosen based on the current maximum stable step sizes $\Delta t_e^*\ ,\ \Delta t_i^*$ to minimize the amount of ion steps: $N_e = \lceil \Delta t_i^* / \Delta t_e^* \rceil\ ,\ \Delta t_i = \Delta t_i^*$.

The steps of the moment-matching Vlasov solver are largely identical to the original ones described in \cite{allmann-rahn-lautenbach-grauer:2022}. However, to improve the fulfillment of Gauss's law, the density correction is applied to the non-Maxwellian part of the distribution function instead of the fluid moments.
Compare Algorithm \ref{alg:timestep} below and in the original publication.

\begin{algorithm}
  Initialize with a half step of the Maxwell solver\\
 \Timestep{}{
  Calculate third moment $\mathcal{Q}^{t}$ from $f^t$\\
  Full Vlasov leapfrog step to advance to $f^{t+1}$\\
  Calculate $\mathcal{Q}^{t+1}$ from $f^{t+1}$\\
  Interpolate to get $\mathcal{Q}^{t+1/2}$\\
  Full Runge-Kutta fluid step (input $\mathcal{Q}$ at appropriate times)\\
  \Momentfitting{}{
    Calculate moments $n_{\text{V}}$, $\mathbf{u}_{\text{V}}$, $\mathcal{P}_{\text{V}}$ from $f$\\
    Calculate ten-moment Maxwellian $f_{\text{M,V}}$ from $n_{\text{V}}$, $\mathbf{u}_{\text{V}}$, $\mathcal{P}_{\text{V}}$\\
    Calculate ten-moment Maxwellian $f_{\text{M,F}}$ from the fluid solver's moments\\
    Exchange ten-moment Maxwellians with density correction: $f = (f - f_{\text{M,V}})\,(n_{\text{F}}/n_{\text{V}}) + f_{\text{M,F}}$\\
    Limit $f$
  }
  Full step of the Maxwell solver\\
 }
 \caption{Time stepping of the moment matching Vlasov-Maxwell solver as it is implemented}
 \label{alg:timestep}
\end{algorithm}

  \subsection{Spatial Coupling}
  \label{sec:coupling}

The spatial coupling of different plasma models can significantly
reduce the computational time needed for a simulation by applying
expensive kinetic models only where kinetic effects are assumed
to be important. In many scenarios, large parts of the physical
domain can appropriately be treated with much cheaper fluid models.

In \muphyii{}, the whole domain is divided into multiple subdomains
that can be computed independently (see the parallelization
information in Sec.~\ref{sec:code_design}). This fits seamlessly
with the spatial coupling multiphysics approach. Each subdomain
can contain different plasma models, and the model coupling at the subdomain
borders is incorporated into the boundary information exchange process.

\begin{figure}[h]
    \centering \includegraphics[width=0.95\textwidth]{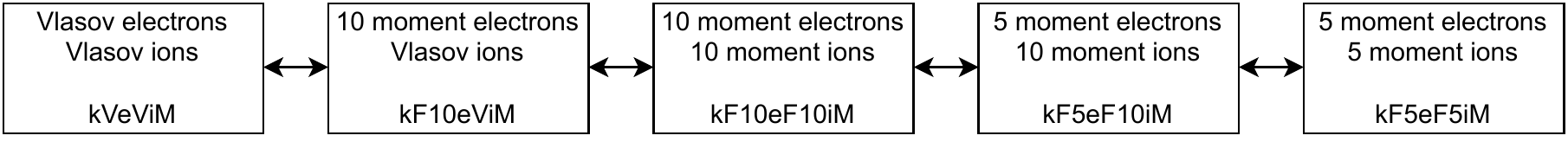}
    \caption{Hierarchy of available models (including the \texttt{enum}s used in \muphyii{}.}
    \label{fig:couplinghierarchy}
\end{figure}

The regions where certain models are used can be static, or they
can be dynamic, i.e.~adapt to the problem over the course of the simulation. For the
latter case, criteria are defined to decide which models are to be used
based on local plasma quantities. A hierarchy of models that are
available for coupling is defined. These models are, starting from
physically complete towards more and more reduced:
full Vlasov $\rightarrow$ hybrid Vlasov ions, ten-moment fluid electrons $\rightarrow$
full ten-moment fluid $\rightarrow$ ten-moment fluid ions, five-moment fluid electrons
$\rightarrow$ full five-moment fluid (see figure \ref{fig:couplinghierarchy}).
We allow coupling only between
neighboring models in this hierarchy to keep the change
in physics at the model boundaries as small as possible. This restriction
is supposed to improve physical accuracy and lead to smooth coupling
borders. In the boundary exchange between different models, 
the more reduced data is sent to minimize the communication
volume. Before or after sending the boundary
data via MPI, it is converted into the appropriate model.

There are two possible model conversions in the hierarchy: Between
the five and ten-moment fluid models or between
the ten-moment fluid and the Vlasov model. We will
discuss both scenarios, starting with the former.
Computing a reduced model from the information of a more
complete model is typically straightforward. Generating five-moment
data out of ten-moment data is done by simply taking the trace
of the ten-moment model's pressure tensor, $\tr(\tensor P)$,
to get the five-moment's scalar pressure.
The other way around is more difficult because there are
degrees of freedom -- an infinite amount of different pressure tensors
can lead to the same scalar pressure. Since in coupled simulations,
we typically use the five-moment model only where the assumption of
isotropic (scalar) pressure is appropriate, we choose the most simple
solution. The ten-moment boundary data is generated out of the
five-moment data by prescribing isotropy, i.e.~setting $\tensor P = p/d\, \tensor I$,
where $\mathrm{P}$ is the pressure tensor, $p$ is the scalar pressure,
and $\tensor I$ the identity matrix.
An alternative implementation available in \muphyii{} extends the pressure tensor
from the ten-moment solution at the coupling border into the boundary cells and
multiplies the pressure tensor by a scalar factor such that
it matches the scalar pressure provided by the five-moment model.

The other model transition is between the Vlasov and the ten-moment model.
Again, it is straightforward to compute the ten-moment data out of
the distribution function boundary data provided by the Vlasov model,
and again, there are infinite possibilities to construct distribution
functions that match the ten-moment boundary data. The first
guess is to use a ten-moment Maxwellian distribution, which takes the temperature
tensor into account (a multivariate normal distribution), as this
is the maximum entropy solution. This approach already leads to good
results. For an even smoother transition, we extend the Vlasov data
at the coupling border into the boundary cells and adjust it to match
the fluid boundary data. This method was introduced in Rieke et al.\ \cite{rieke-trost-grauer:2015}
and Trost et al.\ \cite{trost-lautenbach-grauer:2017}, however
the way of adjusting the distribution function is now done by exchanging
Maxwellians \cite{allmann-rahn-lautenbach-grauer:2022}. We remove the
ten-moment Maxwellian part from the extended Vlasov data and add the result,
multiplied by a factor $\sigma_i$, to the Maxwellian computed from the fluid boundary data.
The factor is chosen to decrease linearly with distance to the coupling
border. In our case of two boundary cells, it is $\sigma_1 = 2/3$ and $\sigma_2 = 1/3$
in the first and second boundary cells, respectively.

  \subsection{Hybrid Ten-Moment-Vlasov Model with Electron Subcycling}
  \label{sec:electron_subcycling}

The hierarchy of models available in \muphyii{} includes a fully
kinetic Vlasov model, fluid models, and a hybrid model. While the
former have been discussed in earlier papers (e.g.\ \cite{allmann-rahn-lautenbach-grauer:2022,allmann-rahn-lautenbach-grauer-etal:2021}),
the latter was used in a simple version in Lautenbach and Grauer (2018) \cite{lautenbach-grauer:2018},
but was not described in detail. Recently, there have been significant additions such
as electron subcycling. The novel hybrid approach that we take, using ten-moment fluid electrons,
Vlasov ions, and full Maxwell's equations, shall be introduced in this section.

Hybrid-kinetic plasma models where the ions are treated kinetically
and the electrons are treated as a fluid have seen much success. They have been
born out of the observation that the ions are more important
for the plasma dynamics in many phenomena while, at the same time, the electrons
are dramatically more expensive to model. Thus, in many cases, it is a good strategy
to approximate the electrons as a fluid so that the kinetic equations need only be
solved on ion time and spatial scales. Traditionally, the electron dynamics
and electromagnetic field equations are approximated via a generalized Ohm's law.
Often, electrons are considered to be a massless fluid with equations of state
for the pressure, typically in the isothermal limit.

There are hybrid-Vlasov codes that solve the ion Vlasov equation on a
phase-space grid, e.g.\ HVM \cite{valentini-travincek-califano-etal:2007,valentini-servidio-perrone-etal:2014}
and Vlasiator \cite{alfthan-pokhotelov-kempf-etal:2014,palmroth-ganse-pfau-kempf-etal:2018}.
On the other hand, there are hybrid-PIC codes that utilize the particle-in-cell
method for the ions, e.g.\ H3D \cite{karimabadi-roytershteyn-vu-etal:2014}
and CAMELIA \cite{franci-hellinger-guarrasi-etal:2018}.
Recently, there have been efforts to include pressure anisotropy into
the fluid electron approximation for H3D \cite{le-daughton-karimabadi-etal:2016}
and HVM (then called HVLF) \cite{finelli-cerri-califano-etal:2021}.

Here, we go even further and employ a complete ten-moment fluid model for
the electrons and use full Maxwell's equations without approximation. With
the help of the electron subcycling and the Maxwell subcycling discussed in
Sec.~\ref{sec:time_stepping} it is still possible to solve the ion Vlasov
equation only on ion time scales. From a physical point of view,
a large ion time step is justified. Numerically, the ions are easier to
handle with respect to their time step because they are not accelerated as strongly,
and the numeric time step restrictions are lower.
The drastic saving in computational
time compared to full Vlasov simulations does not only come from the
larger time step. Since the electrons are much more accelerated than the ions,
especially towards realistic mass ratios, it is necessary to resolve
the large electron velocity space with a high number of cells (mostly due
to numerics). In contrast,
a significantly lower resolution is sufficient for the slower ions, particularly if
the moment matching method is used \cite{allmann-rahn-lautenbach-grauer:2022}.
These two factors together make the hybrid ten-moment-Vlasov model extremely
performant while it still includes the complete ion kinetics, represents electron inertia and light
waves self-consistently, and includes the full electron pressure tensor and a
Landau damping heat flux approximation.
In the past, hybrid-PIC methods were often considered to be more performant
than hybrid-Vlasov methods due to the large velocity space resolutions
necessary for the Vlasov phase-space grid. However, using the matched-moments
Vlasov method, we are able to reduce the ion velocity resolution without loss 
of fidelity and approach a number of freedoms comparable to those used with 
PIC kinetic solvers. We still avoid numerical issues of hybrid-PIC methods 
\cite{stanier-chacon-le:2020}, and discrete particle noise.

\section{Benchmark Simulations}
\label{sec:benchmark_simulations}

In this section, we want to show the features of
\muphyii{} that have not yet been discussed in earlier
publications, which means those that go beyond the
fully kinetic and fully fluid cases.
On the one hand, this is the coupling of different
plasma models, which -- compared to the previous \muphy{} code -- greatly
benefits from the energy-conserving Vlasov solver and the improved
heat flux closure as well as from the technological advancements.
On the other hand, we test the physical expressiveness of the hybrid ten-moment-Vlasov
model, which benefits from the solver improvements as well, but also from the
newly developed electron subcycling approach.

  \subsection{Static Coupling}
  \label{sec:static_coupling}

\begin{figure}
\centering
\includegraphics[width=0.85\textwidth]{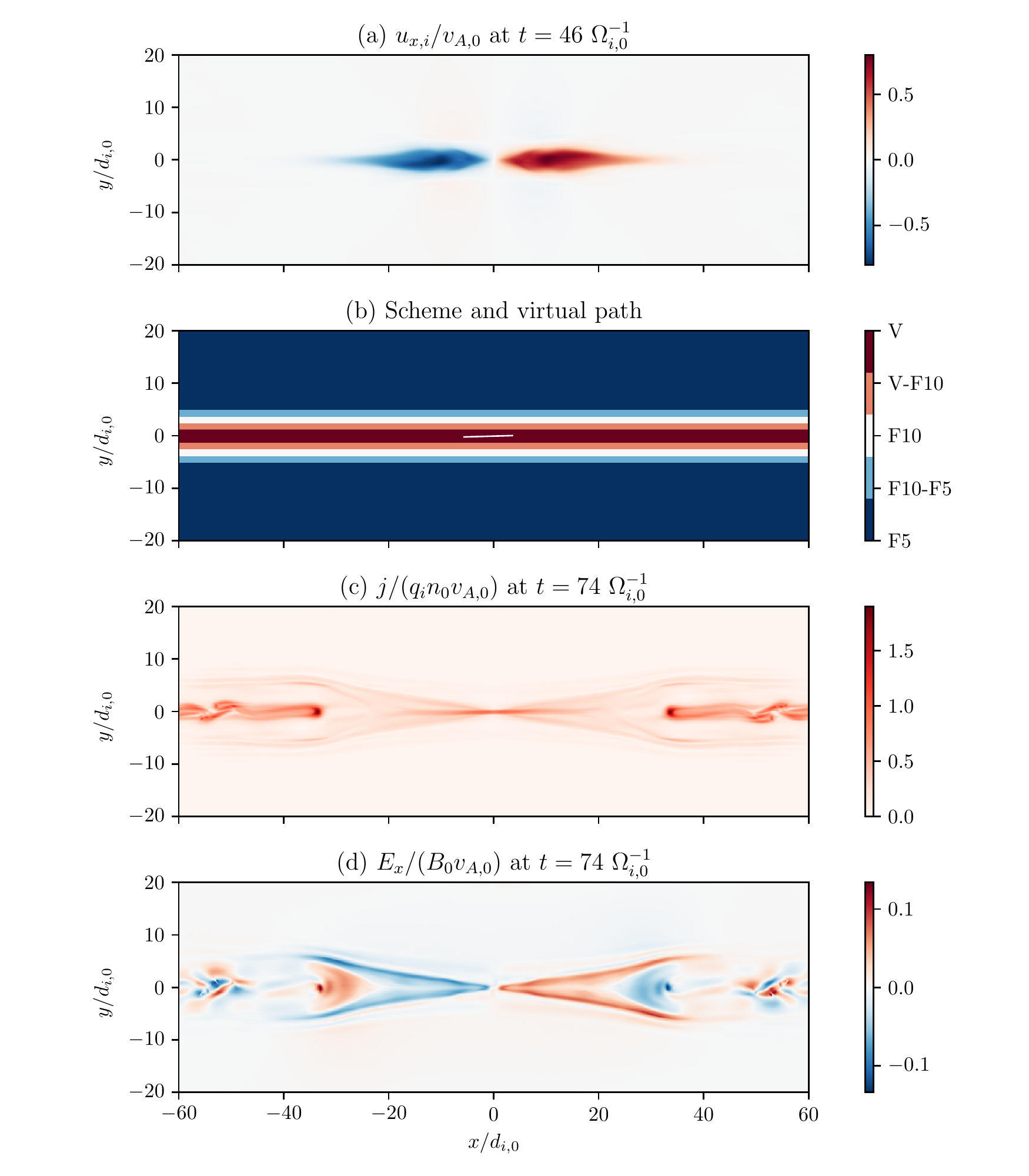}
\caption{Statically coupled simulation of magnetotail
  reconnection. (a) Ion outflow velocity, (b) locations where the
  different numerical schemes are used and the virtual path as
  a white line, (c) absolute value of the current density,
  (d) $x$-component of the electric field.}
\label{fig:magnetotail_ux_coupled}\end{figure}

An ideal setup for testing the spatial coupling is given by the
magnetotail reconnection event measured by MMS on 11th July 2017 at
22:34 UT \cite{torbert-burch-phan-etal:2018} with the initial
conditions derived in Nakamura et al.\ (2018)
\cite{nakamura-genestreti-liu-etal:2018} and Genestreti et al.\ (2018)
\cite{genestreti-nakamura-nakamura-etal:2018}.  This is a good
benchmark case firstly because magnetic
reconnection is an important yet very complex phenomenon
that requires a good representation of kinetic
effects. Secondly, for this particular case,
  spacecraft measurements and highly accurate fully kinetic PIC
\cite{nakamura-genestreti-liu-etal:2018} and Vlasov
\cite{allmann-rahn-lautenbach-grauer:2022} simulations are available
for comparison.

The initial conditions are similar to a Harris equilibrium.  The
initial particle density is given by ${n_s = n_{0}
  \sech^{2}(y/\lambda) + n_{b}}$ and the magnetic field by ${B_{x} =
  \tanh(y/\lambda) B_0 + \delta B_{x}}$, ${B_{y} = \delta B_{y}}$ and
${B_{z} = -B_g}$. The background density is ${n_b = n_0/3}$, the
half-thickness of the current sheet ${\lambda = 0.6\,d_{i,0}}$ and the
guide field ${B_g = 0.03\,B_0}$.  The background particles (those
related to density $n_{b}$) are initially static and have temperatures
${T_{bg,s} = T_{0,s}/3}$ whereas the sheet particles have temperatures
according to ${n_{0} k_{B} (T_{0,e}+T_{0,i}) = B_{0}^{2} / (2
  \mu_{0})}$, ${T_{0,i}/T_{0,e} = 3}$ and carry the current.  The
domain goes from $-L_x/2$ to $L_x/2$ in $x$-direction and $-L_y/2$ to
$L_y/2$ in $y$-direction where ${L_x = 120\,d_{i,0}}$ and ${L_y =
  40\,d_{i,0}}$.  Here, reconnection is initiated by a small Gaussian
perturbation ${\delta B_{x} = -\xi \,(2 y/\lambda^2)\,\exp
  \left(-(x/(a \lambda))^2 \right) \exp\left(-(y/ \lambda)^2\right)
  B_0}$ and ${\delta B_{y} = \xi \,(2 x/(a \lambda)^2)\,\exp
  \left(-(x/(a \lambda))^2 \right) \exp\left(-(y/ \lambda)^2\right)
  B_0}$ where the magnitude of the perturbation is chosen as ${\xi =
  0.025}$ and ${a=L_x/L_y}$.  Electron velocity space ranges from
${-20\,v_{A,0}}$ to ${20\,v_{A,0}}$ and ion velocity space from
${-5\,v_{A,0}}$ to ${5\,v_{A,0}}$.  The ion-electron mass ratio was
set to ${m_i/m_e = 100}$ and the speed of light to ${c =
  20\,v_{A,0}}$. The resolution is ${1536 \times 512}$ cells in
position space times $36^3$ cells in velocity space in the
case of the Vlasov models.  The ten-moment fluid equations
utilize the temperature gradient heat flux closure
\cite{allmann-rahn-lautenbach-grauer:2022} at 16 subcycles.

\begin{figure}
\centering
\includegraphics[width=\textwidth]{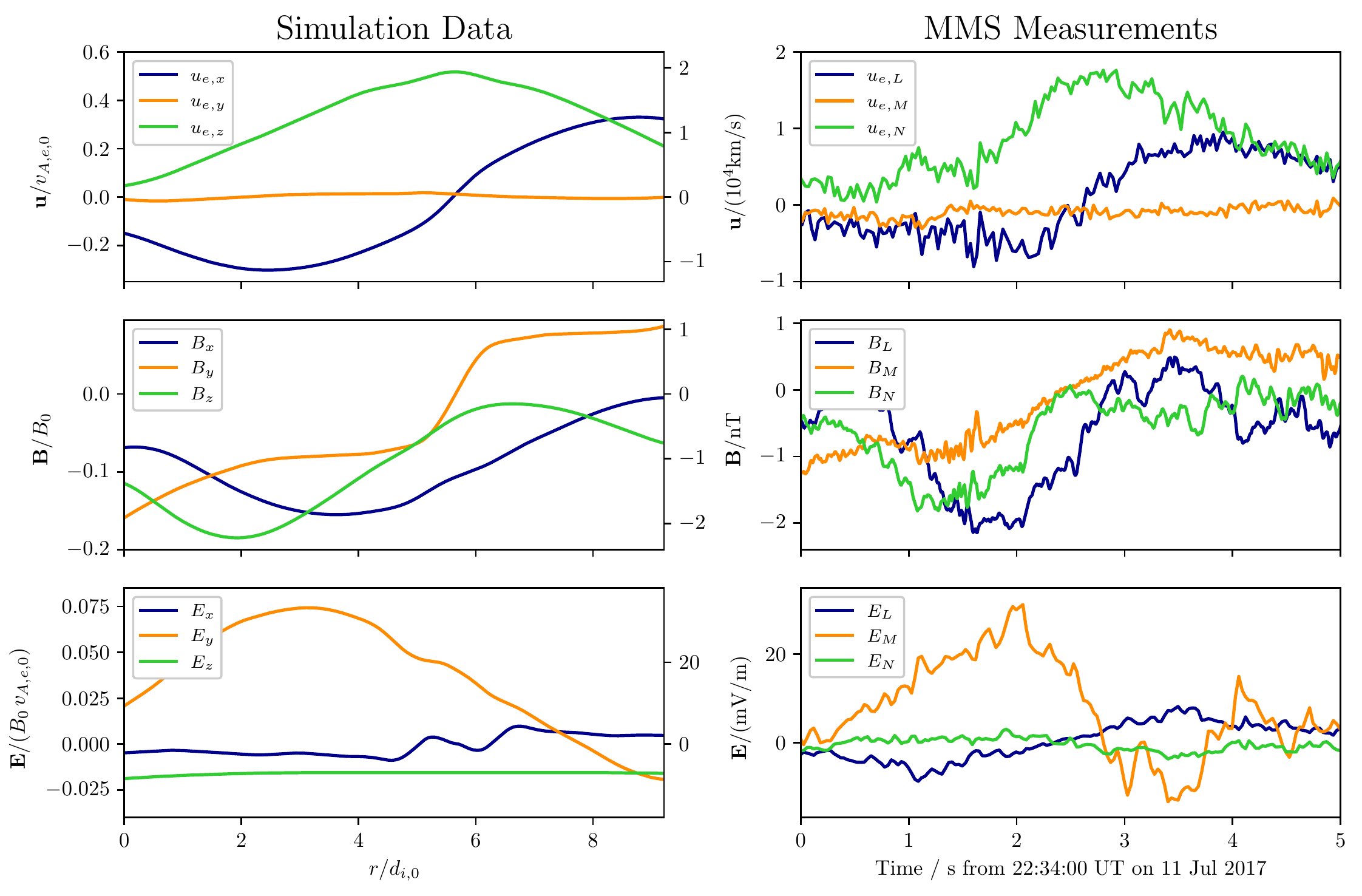}
\caption{Comparison between MMS measurements (right) and the
  statically coupled simulation (left). Simulation data was taken
  along the virtual path shown in
  Fig.~\ref{fig:magnetotail_ux_coupled} at $t=46\,\Omega_{i,0}^{-1}$.}
\label{fig:magnetotail_mms_data_coupled}\end{figure}

An overview of the simulation is given in
Fig.~\ref{fig:magnetotail_ux_coupled}. The structure of the ion
outflow velocity (Fig.~\ref{fig:magnetotail_ux_coupled}a) matches
precisely the one observed in the fully kinetic case (see e.g.\ the
snapshots shown in \cite{genestreti-nakamura-nakamura-etal:2018} and
\cite{nakamura-genestreti-liu-etal:2018}).  As shown in
Fig.~\ref{fig:magnetotail_ux_coupled}b, only a very small
part of the domain is computed with the fully kinetic model, namely
$1/16$, and the largest part is treated with the cheap
five-moment fluid model. Nevertheless, important kinetic
features are captured. In
Fig.~\ref{fig:magnetotail_ux_coupled}c, the absolute value
of the current density is shown at a later time. The elongated current
sheet that is characteristic of kinetic reconnection with
small to moderate guide fields \cite{le-egedal-ohia-etal:2013} is
clearly visible together with an instability in the outflow (possibly
the firehose instability). To demonstrate the smoothness of
the coupling, the $x$-component of the electric field is given in
Fig.~\ref{fig:magnetotail_ux_coupled}d. This is one of the most
sensitive quantities and still shows no signs of irregularities at the
coupling borders, although complex features evolve beyond
the interfaces between the different models.

\begin{figure}
\centering
\includegraphics[width=0.85\textwidth]{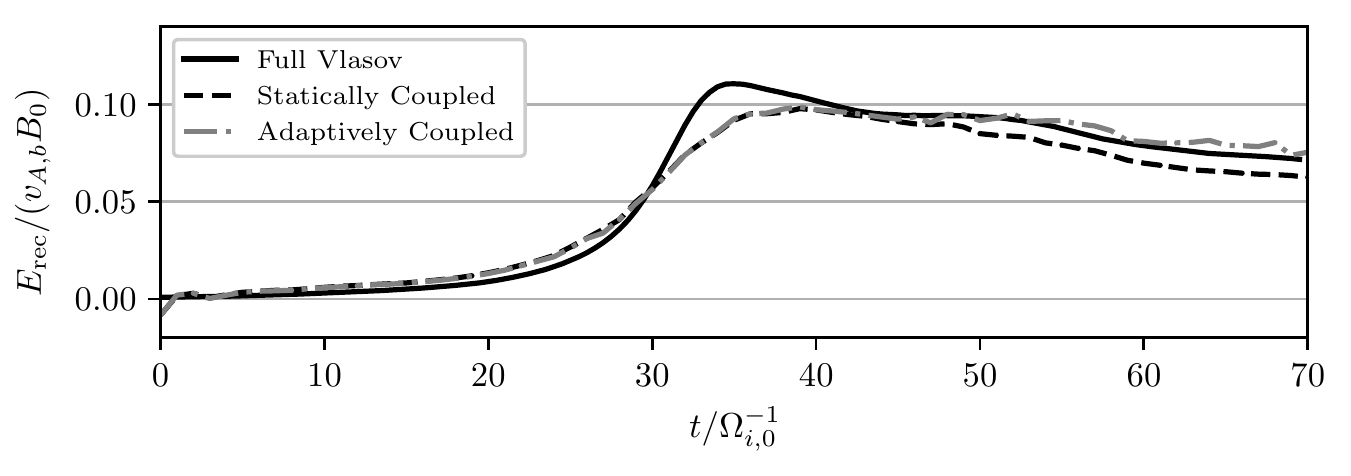}
\caption{Normalized reconnection rate in the statically and
  dynamically coupled simulations compared to the full Vlasov
  simulation from \cite{allmann-rahn-lautenbach-grauer:2022} (the
  latter shifted by $-30\,\Omega_{i,0}^{-1}$ to account for the lower
  initial perturbation used there).}
\label{fig:reconnection_rate_magentotail_comparison}\end{figure}

The coupled simulation compares well to measurements of the MMS
spacecraft of the reconnection event, as shown in
Fig.~\ref{fig:magnetotail_mms_data_coupled}. The simulation data was
taken along the virtual path that is represented by the white line in
Fig.~\ref{fig:magnetotail_ux_coupled}b. While the agreement between
simulation and measurements does not quite reach the quality of fully
kinetic simulations
\cite{nakamura-genestreti-liu-etal:2018,allmann-rahn-lautenbach-grauer:2022}
(which is expected) it does come close. It is clear that the use of
the Vlasov model only in the center of the domain cannot represent all
kinetic effects precisely. In
\cite{allmann-rahn-lautenbach-grauer:2022}, the electron
heat flux was shown to be the strongest at the separatrix
borders, which is absolutely not captured by the
five-moment model used there. In the future, a
better distribution of the models in the domain could further improve
the results. Nevertheless, the result is very encouraging,
considering that the kinetic region is so small.

Similar conclusions can be drawn from a comparison of the normalized
reconnection rate between the coupled simulation and the fully kinetic
Vlasov simulation from \cite{allmann-rahn-lautenbach-grauer:2022}.
The reconnection rate peaks around the typical value of 0.1
\cite{cassak-liu-shay:2017} and is overall quite similar between the
coupled and the fully kinetic simulation. The coupled result is also
comparable to the reconnection rate obtained in
\cite{nakamura-genestreti-liu-etal:2018}, which matches the
full Vlasov case.

  \subsection{Dynamic Coupling}
  \label{sec:dynamic_coupling}
\begin{figure}
\centering \includegraphics[width=0.85\textwidth]{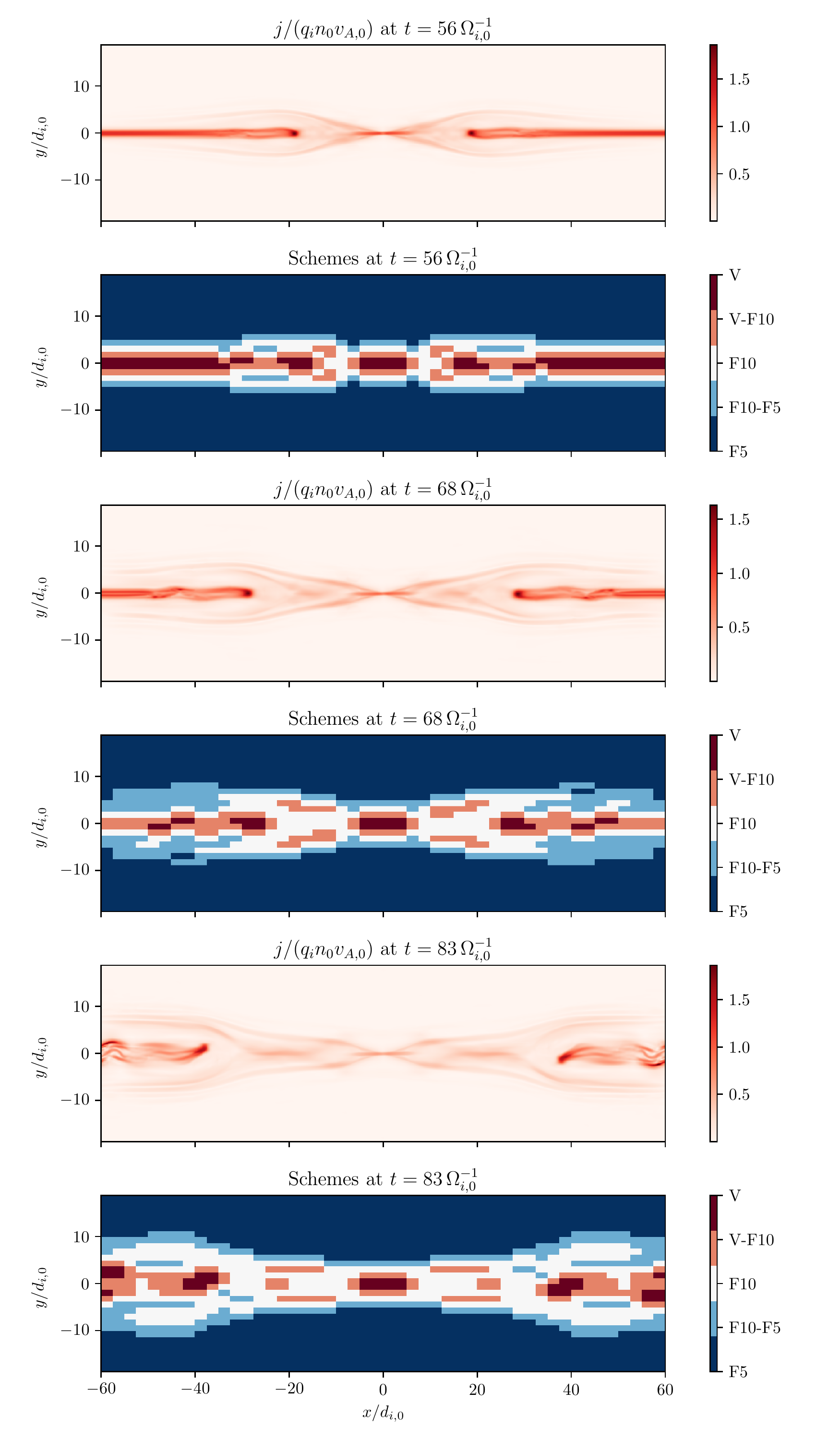}
\caption{Overview of the dynamically coupled simulation of magnetotail reconnection:
Absolute value of the current density and distribution of the different physical
schemes at three different times.}
\label{fig:magnetotail_j_scheme_coupled_adap}\end{figure}

\begin{table}
\centering
\small
\begin{tabular}{p{0.1\textwidth} p{0.38\textwidth} p{0.42\textwidth}}
\toprule
Scheme & Description & Criterion \\
\midrule
V       & Vlasov ions, Vlasov electrons   & $j > 1.00\ n_0 v_{A,0}\ \ \lor\ \ u_e > 4.0\ v_{A,0}$\\
V-F10   & Vlasov ions, 10 mom electrons   & $j > 0.50\ n_0 v_{A,0}\ \ \lor\ \ u_e > 2.0\ v_{A,0}$\\
F10     & 10 mom ions, 10 mom electrons   & $j > 0.20\ n_0 v_{A,0}\ \ \lor\ \ u_e > 1.0\ v_{A,0}$\\
F10-F5  & 10 mom ions, 5 mom electrons    & $j > 0.05\ n_0 v_{A,0}\ \ \lor\ \ u_e > 0.5\ v_{A,0}$\\
F5      & 5 mom ions, 5 mom electrons     & else.\\
\bottomrule
\end{tabular}
\caption{Criteria used for choosing the plasma models in the dynamically coupled simulation.}
\label{tab:criterion}\end{table}

Selecting static regions where the different plasma models are used
is not the optimal approach in most cases because the areas where certain
physical models are required change over time. One example is the electron heat
flux at the separatrix borders in the later simulation time. Therefore, we performed
a simulation with dynamic coupling where a criterion based on local plasma parameters
determines which physical model is used. Here, we choose a heuristic criterion that
is tuned for magnetic reconnection and takes the absolute value of the current density
as well as the out-of-plane electron velocity into account. A similar approach was
used in \cite{lautenbach-grauer:2018}. The thresholds for the different plasma
models are summarized in Table~\ref{tab:criterion}. The model regions are updated
every $5\,\Omega_{i,0}^{-1}$ and the domain is decomposed into $48 \times 32$ blocks
in the $x-y$ plane.

Dynamic coupling comes with challenges. First, the chosen criterion is of course not
optimal, and in the future, a more appropriate criterion needs to be found. Second and
more important, the substitution of numerical schemes can be associated with rather large
inhomogeneities at the interfaces to the neighbor schemes. The reason is that so
far, no process comparable to the spatial Vlasov/ten-moment coupling (see Sec.~\ref{sec:coupling})
has been implemented for the scheme substitution. For example, if a Vlasov model is
coupled to a ten-moment model, then the Vlasov region is extrapolated into the
Maxwellian ten-moment region to improve coupling smoothness. However, if
a ten-moment model region is substituted by a Vlasov model, it starts with a Maxwellian
distribution function and there is no mechanism to improve the smoothness at
the interface. Therefore, the simulation in this section should be seen
as a proof of concept and a demonstration of \muphyii{}'s features, while a solution
to this problem is left for future work.

The simulation state and the scheme distribution are shown at
different times in Fig.~\ref{fig:magnetotail_j_scheme_coupled_adap}.
The reconnection process is represented well, and the
schemes adapt to the geometry of the reconnection current sheet. Despite the
regularly changing distribution of schemes within the domain, and the complex
spatial features, there are no noteworthy artifacts at coupling borders, and
the overall picture is very smooth. Important regions
at the separatrix boundary, and generally the whole separatrix, are treated with
the more advanced models as desired. The evolution of the
reconnection rate is shown in Fig.~\ref{fig:reconnection_rate_magentotail_comparison}
and is close to the fully kinetic case.

We chose the criterion in a way that the fully
kinetic regions are smaller than physically appropriate to show the
numerical robustness of the method. Thus, the kinetic region around the X-line is
not large enough to capture the current sheet extension. Of course, there is 
generally a lot of room for improvement concerning the criterion in the future.
Ideally, much more kinetic physics could be captured at the same computational
cost with an improved distribution of the physical schemes.

  \subsection{Hybrid Ten-Moment-Vlasov}
  \label{sec:hybrid}

In Sec.~\ref{sec:electron_subcycling}, we discuss the high physical expressiveness that is possible
using a ten-moment model for the electrons and full Maxwell's equations in a hybrid fluid electron,
Vlasov ion approach. For a benchmark simulation, we choose the initial conditions from
Finelli et al.\ (2021) \cite{finelli-cerri-califano-etal:2021} where a Landau fluid electron, Vlasov ion model is introduced
that is, as far as the underlying ideas are concerned, quite similar to our hybrid model. However, there are
important differences regarding both numerics and physics that we will discuss.

The physical configuration is, as before, magnetic reconnection in a Harris sheet. Thus, the spatial
profiles of the initial particle distribution and magnetic field are the same as described
in Sec.~\ref{sec:static_coupling}, but the plasma parameters are very different:
The magnetic guide field ${B_g = 0.25\,B_0}$ is much larger and the background density ${n_b = n_0}$ as well.
The current sheet's half-thickness is chosen as ${\lambda = 0.85\,d_{i,0}}$,
ion-electron mass ratio as ${m_i/m_e = 100}$ and temperature ratio as ${T_{0,i}/T_{0,e} = 4}$
with ${T_{bg,s} = T_{0,s}}$. Speed of light is ${c = 20\,v_{A,0}}$.
We choose the same initial perturbation as in the previous sections, now with ${\xi = 0.02}$.
The spatial domain goes from $-L_x/2$ to $L_x/2$ in $x$-direction and $-L_y/2$ to $L_y/2$
in $y$-direction where ${L_x = 24 \pi\,d_{i,0}}$ and ${L_y = 12 \pi\,d_{i,0}}$.
It is resolved by ${1024 \times 512}$ cells. Ion velocity space goes from ${-4\,v_{A,0}}$ to ${4\,v_{A,0}}$
which we intentionally resolve very coarsely using only $16^3$ cells.
Differences to the setup in \cite{finelli-cerri-califano-etal:2021} are, firstly, that we use
a different initial perturbation, as the exact form of the perturbation was not specified there.
Secondly, we choose a \textit{single} Harris sheet configuration contrary to the \textit{double} Harris sheet setup in \cite{finelli-cerri-califano-etal:2021},
which was likely chosen to simplify boundary conditions for the semi-implicit PIC code used there.
Since \muphyii{} does not have such limitations, we prefer the standard single current sheet.

\begin{figure}
\centering \includegraphics[width=0.85\textwidth]{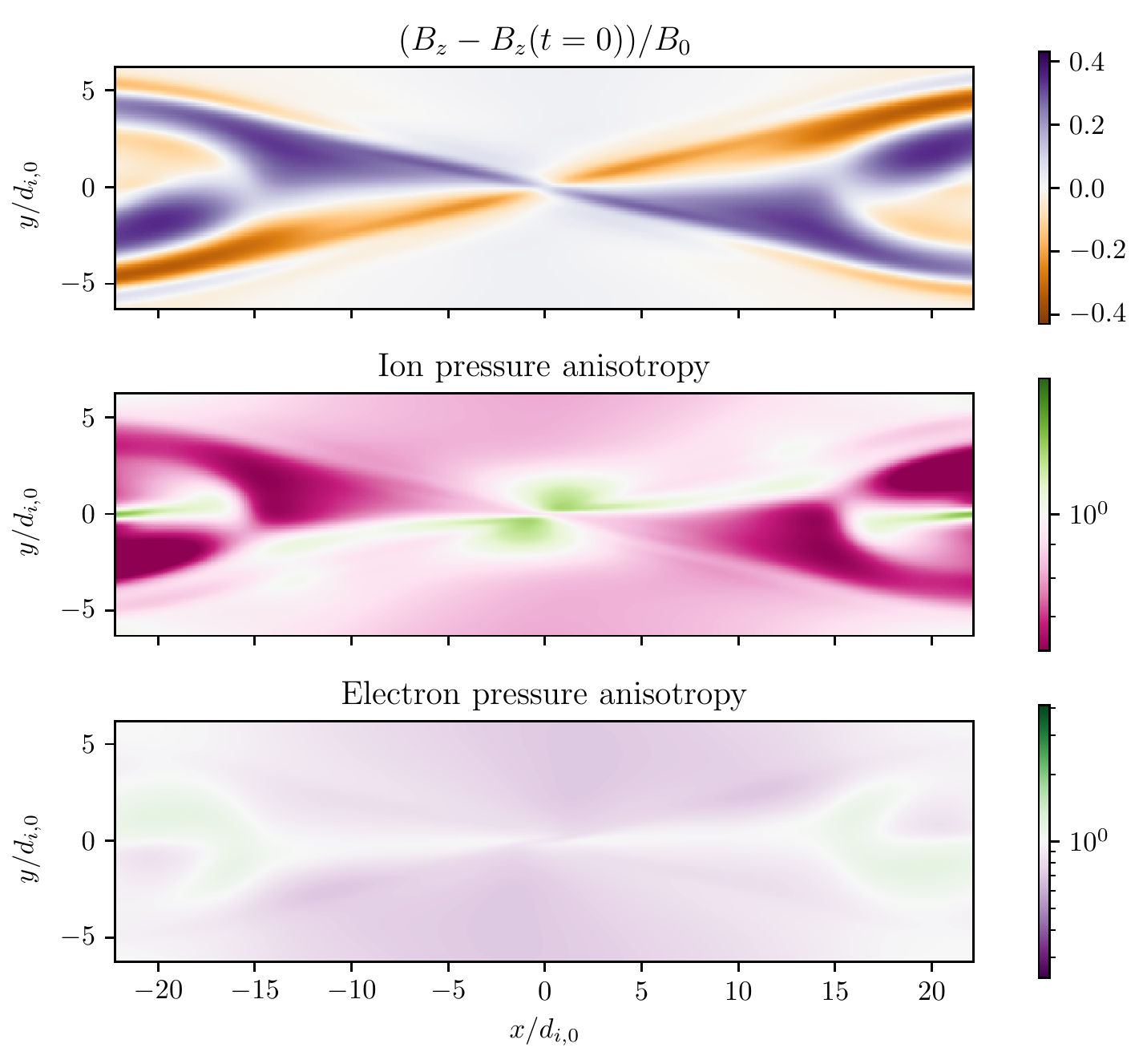}
\caption{Hybrid 10-moment-Vlasov simulation of the reconnection setup from \cite{finelli-cerri-califano-etal:2021},
state at $t=180\,\Omega_{i,0}^{-1}$. Shown are the deviation of the magnetic guide field from its initial
value (top) and the species pressure anisotropies (middle, bottom).}
\label{fig:finelli_bz_anisotropy}\end{figure}

Of course, the chosen mass ratio is not one where the hybrid model can exploit its full potential, and a far
higher spatial resolution is possible as well. However, we want to use the same parameters
as in \cite{finelli-cerri-califano-etal:2021} to make a direct comparison possible with the results therein and
the respective models, which
include a standard hybrid simulation (code HVM), the mentioned hybrid Landau fluid simulation (code HVLF, an extension of HVM) and a
fully kinetic simulation using a semi-implicit PIC model (code iPIC3D). As opposed to these, the \muphyii{} simulation
that we present in this section has an exceptionally low computational cost, as we will point out.
We want to elaborate on the physical differences between these models briefly. The HVM code uses the most common
hybrid model with a quasi-neutral electron fluid, fully isotropic electron pressure, and a generalized Ohm's law.
The HVLF model also makes the quasi-neutral and generalized Ohm's law approximations (including electron inertia) but evolves the gyrotropic pressure
(i.e.\ the components parallel and perpendicular to the magnetic field) with a Landau fluid closure for the gyrotropic
heat flux. In contrast, the \muphyii{} hybrid model uses the ten-moment fluid equations for the electrons, which evolve
the full pressure tensor with the temperature gradient heat flux closure to model Landau damping, and full Maxwell's equations. Therefore, it
self-consistently captures -- alongside many other physical effects -- electron inertia, charge separating plasma waves and light
waves. The implemented heat flux closure is a general-purpose closure for collisionless wave damping, whereas the closure
implemented in HVLF also considers the direction imposed by the magnetic field. Therefore, it is expected that the HVLF
fluid closure is better suited for guide field reconnection (which is the setup in this section) but cannot be
utilized for reconnection where the guide field is zero or close to zero (as in the previous sections).
The iPIC3D code uses a fully kinetic model for both electrons and ions, as well as full Maxwell's equations so that all physics
can be represented and no approximation for heat flux is needed.
At the given resolution, the implicitness of iPIC3D removes charge-separating waves and drives
towards quasi-neutrality \cite{finelli-cerri-califano-etal:2021}.

In Fig.~\ref{fig:finelli_bz_anisotropy} three quantities are shown that are relevant in estimating the model's
capabilities of capturing anisotropy: The magnetic field in guide field direction (difference from initial value) $\tilde{B_z} = B_z(t)-B_z(t=0)$, and
the pressure anisotropies ${P_\bot}/{P_\parallel}$ for ions and electrons given by
\begin{equation*}
\frac{P_\perp}{P_\parallel} = \frac{\sum_{i,j \in \{x,y,z\}}P_{ij}(\delta_{ij}-b_ib_j)/2}{\sum_{i,j \in \{x,y,z\}}P_{ij}b_ib_j}\quad \text{where } b_i = B_i/\abs{\vec B}, P_{ji}=P_{ij}
\end{equation*}

We compare the hybrid ten-moment electron, Vlasov ion results from \muphyii{} with Figures 3 and 4 of \cite{finelli-cerri-califano-etal:2021}, which
represent the simulation data produced by the codes HVM, HVLF, and iPIC3D.
The asymmetry in the spatial structure of $\tilde{B_z} = B_z(t)-B_z(t=0)$ shown in the top panel of Fig.~\ref{fig:finelli_bz_anisotropy}
is caused to a large extent by the electron anisotropy and agrees very nicely with the fully kinetic results from iPIC3D,
in particular around the Alfvénic front in the outflow region, where there are some differences between iPIC3D and the HVLF model.
As expected, around the X-point this asymmetry is better represented in the HVLF simulation, which uses a heat flux
closure that is well-suited for guide field reconnection. The HVM model, on the other hand, produces a nearly symmetrical
result, which is in contrast with the kinetic findings.

The ion pressure anisotropy shows similarities between all models. Interestingly, both the hybrid ten-moment electron \muphyii{}
simulation (middle panel of Fig.~\ref{fig:finelli_bz_anisotropy}) 
and the HVLF simulation have higher values of ion pressure anisotropy ${P_{\bot,i}}/{P_{\parallel,i}}$ in the Alfvénic outflow than the iPIC3D
simulation. Whether this is caused by numerical differences between the Vlasov and the implicit PIC method, by different
physics in the hybrid and the fully kinetic models, or simply by different states of the simulations cannot be settled here.

Last, we want to discuss the electron pressure anisotropy. The picture fits well with what already became clear from the
analysis of the magnetic field in the guide field direction. The overall structure of electron pressure anisotropy in the
hybrid ten-moment electron simulation (bottom panel of Fig.~\ref{fig:finelli_bz_anisotropy}) differs from the iPIC3D result.
The HVLF clearly performs better around the X-point, predicting
lower values of ${P_{\bot,e}}/{P_{\parallel,e}}$, just as the fully kinetic model, and reproduces the elongation of the current sheet
that is typically observed in guide field reconnection. This reverses around the Alfvénic front, where
${P_{\bot,e}}/{P_{\parallel,e}} > 1$ in both the hybrid ten-moment electrons result from \muphyii{} and the
kinetic iPIC3D result, but not in HVLF. Both observations are to be expected, considering that the gyrotropic formulation of the HVLM
electron closure excels in regions with a strong directional magnetic field, while the \muphyii{} electron closure has been developed
from the unmagnetized limit and has its strength in corresponding regions of low magnetization.

To summarize the results, it was shown that the hybrid ten-moment electron, Vlasov ion model performs to expectations in
this guide field reconnection setup, although it was not tuned for the physical configuration in the slightest.
More importantly, the \muphyii{} hybrid simulation (which was run using only eight GPU cards) was computationally
much cheaper than the simulations in \cite{finelli-cerri-califano-etal:2021}.
The primary reason for this is that the matched-moments Vlasov solver can deal with the low velocity space
resolution of $16^3$ cells. For comparison, the resolution used for the HVM and HVLF runs in \cite{finelli-cerri-califano-etal:2021}
is $51^3$ cells for velocity space and would result in 32 times more degrees of freedom, and thus a 32 times higher
computational effort for solving the Vlasov equation, though the actual runtime is not as easily comparable if
the models' numerical schemes and timestep requirements differ.
Still, despite the considerably lower velocity resolution, there
are no major qualitative differences in the ion results between the higher resolved HVM and HVLF runs and the lower
resolved \muphyii{} runs that cannot be attributed to the difference in electron models. That is no surprise either,
as the higher resolution is primarily necessary for numerical
reasons in classic Vlasov methods, and low velocity space resolutions have been used in
PIC codes with much success since the beginning. Since the moment-matching approach is not dependent on numerical schemes of \muphyii{},
and is transferable to other implementations like HVM and HVLF (with appropriate treatment of the second moment), we expect
those models could benefit from its reduced computational footprint as well. A second performance aspect is that the
heat flux closure in the HVLF model utilizes computations in Fourier space, which is much more expensive at large problem sizes
than the local closure used by \muphyii{}.

Through the use of electron and Maxwell subcycling and the {moment-matched} Vlasov solver, the novel hybrid
model implemented in \muphyii{} can represent detailed electron and electromagnetic physics in collisionless
plasmas at very low computational cost.

\section{Conclusions}
\label{sec:conclusions}
In this paper, we have presented the multiphysics framework \muphyii{} for
simulations of collisionless plasmas in space and astrophysical environments.
We interface-couple comparatively cheap fluid solvers with kinetic Vlasov
solvers and employ the moment-matching approach, subcycling of
Maxwell's equations, and electron subcycling. As a result, the simulation of spatial
regions that would not be treatable with Vlasov solvers alone is achieved.
We have also shown that the moment-matching approach benefits the hybrid
Vlasov-Fluid model in particular.  
In conclusion, this framework is an important step towards achieving the goal of global
simulations from kinetic to magnetohydrodynamic (MHD) scales.

It can be even more rewarding to use the coupling
of different plasma models in three-dimensional simulations. In 3D, depending on the
physical problem, the portion of the spatial domain that can be covered by fluid models
may be much larger than in 2D. \muphyii{} has already been used for three-dimensional
ten-moment fluid simulations \cite{allmann-rahn-lautenbach-grauer-etal:2021}, and the
code fully supports coupled simulations in 3D, which we plan to utilize in the future.
There, the matched-moments Vlasov solver will be of great value because it can reduce
the large memory requirements that are inherent to the Vlasov method in three dimensions.

The future path to extending this framework is fairly obvious. To achieve true
multiscale/multiphysics simulations, the coarsest model in our framework, the
five-moment/Maxwell model, needs to be coupled with the MHD equations equipped
with Ohm's law and extended with adaptive mesh refinement (AMR). This is an
ongoing work to combine the multiphysics framework \muphyii{} and our multiscale
framework \racoon{} \cite{dreher-grauer:2005}.

In addition, the framework \muphyii{} allows but also needs further, more
sophisticated implementations of refinement criteria. The design of these
criteria, e.g. based on some form of suitably defined residuum, is a major
mathematical challenge for this kind of simulations, not only in plasma
physics.

\section*{Code Availability}
\label{sec:code_availability}
The \muphyii{} code is available on GitHub (\url{https://github.com/muphy2-framework/muphy2}) and Zenodo (\url{https://zenodo.org/doi/10.5281/zenodo.8061586}).

\section*{Acknowledgements}
We gratefully acknowledge the Gauss Centre for Supercomputing e.V.
(www.gauss-centre.eu) for funding this project by providing computing time
through the John von Neumann Institute for Computing (NIC) on the GCS
Supercomputer JUWELS \cite{jsc-juwels:2021} at Jülich Supercomputing Centre (JSC).
Computations were conducted on JUWELS/JUWELS-booster and on the DaVinci cluster
at TP1 Plasma Research Department. F.A.~was supported by the Helmholtz Association (VH-NG-1239).
M.D. acknowledges funding from the German Science Foundation DFG through the research unit “SNuBIC” (DFG-FOR5409).
R.G. acknowledges support from the German Science Foundation DFG, within the Collaborative Research Center SFB1491 “Cosmic Interacting Matters - From Source to Signal”. 
Thanks to the MMS team for the measurement data available at the
MMS Science Data Center (https://lasp.colorado.edu/mms/sdc/), and to the
developers of the pySPEDAS software (https://github.com/spedas/pyspedas) as well as
the developers of the SpacePy software (https://spacepy.github.io/).

\bibliography{bibliography}

\end{document}